\documentclass[aps,twocolumn,superscriptaddress,showpacs,amsmath,amssymb]{revtex4-1}

\usepackage{graphicx}
\usepackage{dcolumn}
\usepackage{bm}
\usepackage[latin1]{inputenc}

\expandafter\ifx\csname package@font\endcsname\relax\else
 \expandafter\expandafter
 \expandafter\usepackage
 \expandafter\expandafter
 \expandafter{\csname package@font\endcsname}%
\fi
\usepackage{bm}%
\usepackage[colorlinks=true,linkcolor=blue]{hyperref}%
\usepackage[latin1]{inputenc}
\usepackage{graphicx}
\usepackage{dcolumn}
\usepackage{bm}

\begin{document}

\title{Magnetic Dynamics and Spin Freezing in the Ferromagnetic Phase of Ba-Doped Perovskite Cobaltites}

\author{M. Stingaciu}
\affiliation{Laboratory for
Development and Methods, Paul Scherrer Institut, CH-5232 Villigen,
Switzerland} \affiliation{Department of Materials and Environmental Chemistry, Stockholm University S-106 91 Stockholm, Sweden}

\author{H.~Luetkens} \email{{Corresponding author: hubertus.luetkens@psi.ch}} \affiliation{Laboratory for
Muon-Spin Spectroscopy, Paul Scherrer Institut, CH-5232 Villigen
PSI, Switzerland}
\author{Y.G. Pashkevich}
\affiliation{A.A. Galkin Donetsk Phystech NASU, 83114 Donetsk,
Ukraine}

\author{E. Pomjakushina} \affiliation{Laboratory for
Development and Methods, Paul Scherrer Institut, CH-5232 Villigen,
Switzerland}

\author{K. Conder} \affiliation{Laboratory for Development
and Methods, Paul Scherrer Institut, CH-5232 Villigen,
Switzerland}

\author{H.-H.~Klauss}
\affiliation{Institut für Festkörperphysik, TU Dresden, D-01069 Dresden, Germany}


\date{\today}

\begin{abstract}
The magnetic properties of polycrystaline
La$_{0.5}$Ba$_{0.5}$CoO$_{2.5+\delta}$ ($\delta$ = 0.44 and 0.485)
where investigated using muon spin relaxation and bulk
magnetization experiments. Below a Curie temperature of $T_C =
160$~K and 180~K, respectively, a magnetic phase with a
macroscopic ferromagnetic moment forms. $\mu$SR proves that the
full volume inhomogeneously orders at $T_C$ but that there is
microscopically a phase separation into phase volumes possessing a
fully static and a slow magnetic dynamic behavior. A peak in the
dynamic relaxation rate 35~K respectively 55~K below the Curie
temperature indicates the freezing of spin components. The spin
fluctuations are thermally activated with a typical Jahn-Teller
like phonon energy of $E_\mathrm{a} = 340$~K. Together with
co-operative Jahn-Teller distortions appearing below the spin
freezing temperature, this points to a strong magneto elastic
coupling as the cause for the magnetic fluctuations and the spin
freezing.
\end{abstract}



\maketitle

\section{Introduction}

Materials with competing magnetic exchange interactions may adopt
a randomly frozen arrangement of spins at low temperatures know as
spin glass \cite{binder,mydosh}. Even starting from a long range
magnetically ordered state, competing ferromagnetic (FM) and
antiferromagnetic (AFM) interactions can leads to a loss of
collinearity and the formation of a spin glass \cite{brand}. Also
the phase separated state of a partially randomly frozen spin
glass and the collinear magnet can be achieved. The nature of the
spin glass formation wether it is achieved through an equilibrium
phase transition or a more gradual freezing process has been
strongly debated. For the study of spin dynamics, the muon spin
rotation ($\mu^+$ SR) technique has been recognized to be a
powerful probe among other complementary experimental methods. The
possibility of zero-field measurements makes the technique well
suitable for the spin glass systems for which a very small applied
external field can change the internal configuration of the
system.

Cobaltites of the chemical composition RBaCo$_2$O$_{5+\delta}$
have recently attracted great interest due to their intriguing
electrical and magnetic properties
\cite{Taskin05a,Moritomo00,Troyanchuk98b,Maignan99,Akahoshi99,Plakhty05,Pomjakushina06,Wu01,Zhitlukhina07,Coutanceau95,Foerster01,Khomskii04,Fauth02,luetkens}.
On top of orbital/charge ordering, metal-insulator transition and
giant magnetoresistance effects an additional degree of freedom
namely the different possible spin states of Co make these
materials a fascinating area for fundamental science research. The
possibility to adopt different spin states in the cobaltites is
due to the competition of the crystal field splitting (10~Dq) of
the 3d states favoring lower spin state configurations and the
Hund's exchange interaction (J$_H$) that lowers the energy of each
pair of electrons with parallel spin. For example in LaCoO$_3$ the
ground state is a nonmagnetic Co$^{3+}$(S~=0, t$^6_{2g}$$e^0_{g}$)
due to the fact that 10~Dq is slightly larger than J$_H$. A spin
state transition to a higher spin state takes place with
increasing temperature. The mechanism of the transition is still
under debate and different scenarios were proposed. Some authors
concluded a change from low spin to intermediate spin (S~=1,
t$^5_{2g}$$e^1_{g}$) while others suggest the formation of the
high spin state configuration (S~=2, t$^4_{2g}$$e^2_{g}$)
\cite{raccah,senaris,bhide,yamaguchi}. The introduction of
additional charges by e.g. the substitution of La$^{3+}$ by
M$^{2+}$ (M=~Ba, Sr, Ca) in LaCoO$_3$ results in remarkable
changes in their magnetic and electrical properties.

In this study, we have investigated powder samples of
La$_{0.5}$Ba$_{0.5}$CoO$_{2.94}$ and
La$_{0.5}$Ba$_{0.5}$CoO$_{2.985}$. Compared to the charge-balanced
RBaCo$_2$O$_{5.5}$ compound, these samples are hole-doped and
possess a random distribution of La and Ba cations on the R site.
The increased oxygen content introduces a corresponding amount of
Co$^{4+}$ into the undoped system consisting of Co$^{3+}$ only.
For the fully oxidized compound, La$_{0.5}$Ba$_{0.5}$CoO$_{3}$ a
1:1 mixture of Co$^{3+}$ and Co$^{4+}$ is obtained. The physical
properties are strongly influenced by this doping due to competing
magnetic interactions. A predominantly FM order due to the double
exchange interactions between Co$^{3+}$-Co$^{4+}$ compete with the
AFM exchange mediated by the superexchange mechanism. This
competition may lead to the coexistence of FM and glassy behavior
observed e.g. in La$_{0.5}$Sr$_{0.5}$CoO$_3$ \cite{Nam99}. The
A-site ordered compound LaBaCo$_2$O$_6$ possesses a ferromagnetic
transition at T$_\mathrm{C}$ =175~K while in its isochemical
counterpart La$_{0.5}$Ba$_{0.5}$CoO$_3$ the T$_\mathrm{C} =
190$~K. Below around 140~K both La$_{0.5}$Ba$_{0.5}$CoO$_3$ and
LaBaCo$_2$O$_6$ transform to a magnetic glassy state
\cite{nakajima}. At the same temperature Fauth \emph{et al.}
\cite{fauth} observed a structural change from cubic to tetragonal
which has been interpreted as a cooperative Jahn-Teller effect and
a partial d$_{3z^2-r^2}$ type of orbital ordering. This
interpretation is supported by the observed elongation of the
CoO$_6$ octahedra along the direction of the apical oxygen. At the
same temperature a metal to insulator transition is observed.
Since mobile charges are a prerequisite for the FM double exchange
mechanism, a change in the balance of the competing AFM-FM can be
anticipated at this temperature.

In this paper we present a susceptibility and $\mu^{+}$SR study of
La$_{0.5}$Ba$_{0.5}$CoO$_{2.5+\delta}$ for two samples with
different oxygen content, $\delta$ =~0.44 and 0.485. Macroscopic
dc and ac-susceptibility measurements reveal a FM transition at
$T_C=160$~K and 180~K respectively. A clear frequency dependence
of the ac susceptibility is observed below $T_C$, i.e. a magnetic
behavior typically found in systems exhibiting a spin freezing.
$\mu$SR reveals quasi-static disordered magnetic ordering below
$T_C$ with slow magnetic dynamics persisting in 70\% of the sample
volume down to lower temperatures. Far below $T_C$ (35~K and 55~K)
a peak in the dynamic relaxation indicates the freezing of the
fluctuating spin components. The measured magnetic fluctuation
rates can be well described by a thermally activated process with
an activation energy typical for Jahn-Teller like phonons. This
and the previously observed Jahn-Teller distortions in the crystal
lattice at the same temperature indicate that the freezing process
is governed by a magneto elastic coupling.

\section{Experimental}
 La$_{0.5}$Ba$_{0.5}$CoO$_{2.5+\delta}$
($\delta$ = 0.44 and 0.485) samples were synthesized by a standard
solid state reaction. The rare earth oxide, cobalt oxide and
barium carbonate of a minimum purity of 99.99\% were mixed and
annealed at temperatures 900 - 1200~$^\circ$C for 100~h in air,
with several intermediate grindings. The obtained material was
separated into two batches. One of the batches was additionally
annealed in oxygen flow for 20 hours. The oxygen content was
determined with an accuracy of $\pm$0.01 of the oxygen index in
the chemical formula by iodometric titration. Since the synthesis
requires elevated temperatures and high oxygen pressures to obtain
a layered perovskite structure \cite{nakajima}, a slight
oxygen-deficiency remains without a high-pressure oxygen
annealing. We found for the as-prepared sample an oxygen content
of $\delta$~=~0.44 whereas for the oxidized sample
$\delta$~=~0.485. The details of the sample preparation and
determination of the oxygen content can be found in
Ref.~\cite{conder}. From the analysis of our X-ray data using the
Rietveld profile refinements a cubic structure with a space group
Pm$\overline{3}$m was found at room temperature for both sample
batches. The cubic symmetry implies a complete disordering of the
lanthanum and barium ions on the same site of the perovskite. The
bulk magnetic properties have been studied using a Physical
Properties Measurements System (PPMS) in a temperature range $T
=~10 - 250$~K. The $\mu$SR technique utilizes positively charged
$\mu^+$ which are implanted into the sample and thermalize at
interstitial lattice sites, where they act as a volume sensitive
magnetic microprobes. The amplitude of the different $\mu$SR
signals is proportional to the volume fraction of the
corresponding magnetic phase. Therefore $\mu$SR is an ideal tool
to investigate phase separation phenomena in magnetic materials.
Here, only a short description of the used time-differential
method is given while a more elaborate discussion about the
application of $\mu$SR to magnetic systems can be found elsewhere
\cite{Reotier97}. A continuous muon beam with initial polarization
close to 100~\% is implanted into the sample. The muon spin
precesses about possibly existing local magnetic fields
$B_\mathrm{loc}$ with the Larmor frequency $2\pi f = \gamma_\mu
B_\mathrm{loc}$ (muon gyromagnetic ratio $\gamma_\mu=8.531\times
10^8$~rad~s$^{-1}$~T$^{-1}$). With a lifetime of
$\tau_\mu=~2.2$~$\mu$s the muons decay into two neutrinos and a
positron, the latter being predominantly emitted along the
direction of the muon spin at the moment of the decay. Measurement
of both the direction of positron emission as well as the time
between muon implantation and positron detection therefore
provides a sensitive determination of the muon spin polarization
$P(t)$. The time evolution of $P(t)$ depends on the distribution
of internal magnetic fields and their temporal fluctuations. In
this work, the measurements were performed in zero applied field
(ZF-$\mu$SR), which is especially useful when even small external
fields might change the intrinsic magnetic state of the system
under investigation. $P(t)$ is measured via the decay asymmetry
$A(t)$ in forward and backward positron counters at time $t$:
\begin{equation}\label{1}
A(t)= A_0  P(t)
\end{equation}
Here, $A_0$ is the experimentally observable decay asymmetry of the order of 0.27 and
\begin{equation}\label{2}
A(t)=\frac{B(t)- \alpha F(t)}{B(t)+ \alpha F(t)},
\end{equation}
where the $B(t)$ and $F(t)$ are the number of positrons detected
in the backward and forward detectors, respectively, at time $t$
after the arrival of the corresponding muon in the sample. The
$\alpha$ parameter is a calibration constant which usually has a
value close to one and accounts for solid angle and efficiency
difference of the two detectors. In our experiments the
polycrystalline samples were packed into a thin polyethylene foil
and mounted on a fork holder in a helium flow cryostat in order to
ensure that no background signal is measured from the holder and
the packing.

\section{Results and discussion}
\begin{figure}[htbp]
\begin{center}
\center{\includegraphics[width=0.87\columnwidth,angle=0,clip]{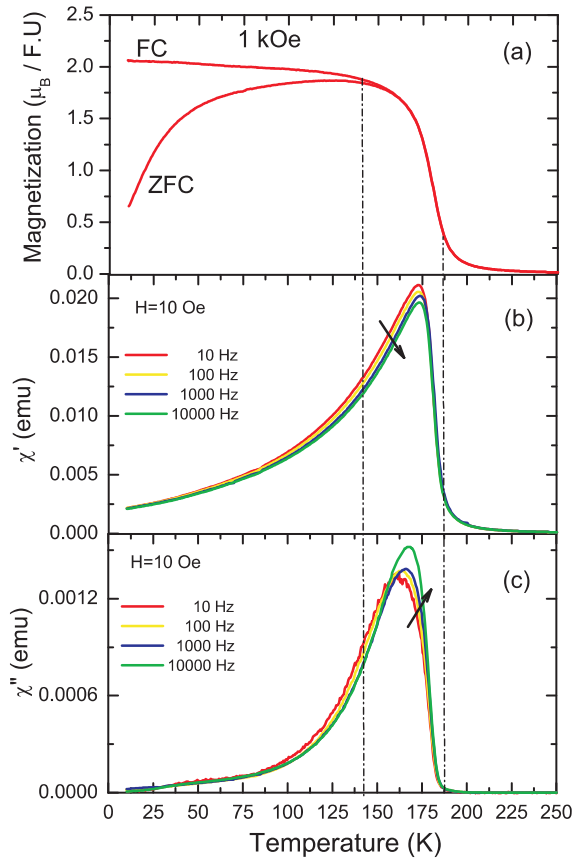}}
\caption{dc and ac-susceptibility measurements performed for
La$_{0.5}$Ba$_{0.5}$CoO$_{2.985}$. (a) The temperature dependence
of ZFC and FC magnetization measurements performed in 0.1~T
external magnetic field. (b) Temperature dependence of the
in-phase $\chi^{'}$ measured with ac-field of 10 Oe for different
frequencies. (c) The corresponding out of phase $\chi^{''}$. The
arrows show the shift of susceptibility with increasing
frequency.} \label{Fig-Magn}
\end{center}
\end{figure}

Fig.~\ref{Fig-Magn}(a) shows dc-magnetization measurements performed
with an applied external field of 0.1~T after zero field cooling
(ZFC) and field cooling (FC) in the temperature range 10-250~K for
La$_{0.5}$Ba$_{0.5}$CoO$_{2.985}$.
For temperatures higher than 180~K the sample is paramagnetic. In
fact at T$_C$ = 180~K a ferromagnetic transition is observed with
a sudden increase of the magnetization. Lowering the temperature
the magnetization exhibits an irreversible magnetic behavior
between ZFC and FC magnetization curves below 140~K. The
saturation moment is close to 2~$\mu_B$/F.U. indicating a mixture
of spin states which are most probably IS/LS of both Co$^{3+}$ and
Co$^{4+}$ ions \cite{fauth,nakajima}. The ac-susceptibility
measurements as a function of temperature are shown in
Fig.~\ref{Fig-Magn}(b) and \ref{Fig-Magn}(c) for the in-phase
$\chi'$ and out of phase $\chi^{''}$ component of the
susceptibility, respectively. The magnetic susceptibility shows a
sharp increase close to the FM transition. At lower temperature,
slightly below T$_c$ the susceptibility measurements show a peak
followed by slow decrease with decreasing temperature. The
observed peak as well as the low temperature tail of $\chi^{'}$
shift to higher temperatures with increasing measuring frequency.
The shift to the higher temperatures with increasing the frequency
is better observed for $\chi^{''}$ in Fig.~\ref{Fig-Magn}(c). The
magnetic behavior of La$_{0.5}$Ba$_{0.5}$CoO$_{2.985}$, i.e. the
dc-magnetization below 140~K, the frequency dependent cusp in the
ac-susceptibility and the small hysteresis, are consistent with a
spin glass and cluster glass like freezing of the magnetic moments
\cite{nam,wu}. The actual temperature where ZFC and FC
magnetization deviate from each other depends on the externally
applied field also in accordance with spin glass like systems.
Interestingly our susceptibility data are in good agreement with
previous measurements on the A-site ordered compound
LaBaCo$_2$O$_6$ \cite{kundu} which indicates that this structural
feature is not of strong importance for the magnetic behavior
studied here.

\begin{figure}[htbb]
\begin{center}
\center{\includegraphics[width=0.87\columnwidth,angle=0,clip]{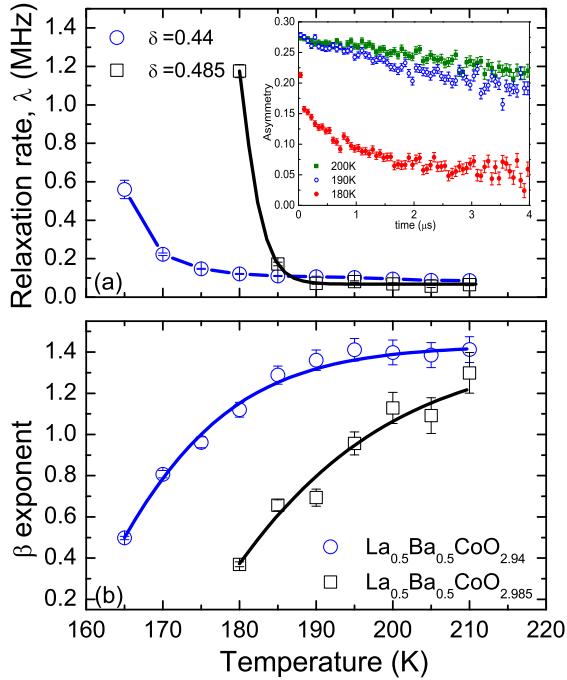}}
\caption{(a) The temperature dependence of the dynamic relaxation
rate as defined by Eq.~\ref{P(t)aboveTc}. Lines are guide to the
eye, only. The inset shows ZF-$\mu$SR raw data for $T=200$, 190
and 180~K.~(b)~The stretched exponential exponent $\beta$ as a
function of temperature for T $>$ T$_c$.}\label{Fig-mSR-highT}
\end{center}
\end{figure}

In order to get a better understanding of the magnetic order and
to probe the spin dynamics in
La$_{0.5}$Ba$_{0.5}$CoO$_{2.5+\delta}$, ZF-$\mu^{+}$SR was used.
The technique can give a direct measure of the internal fields in
bulk samples. The sensitivity of the technique to fluctuations of
the internal field gives important information about the dynamics
of the magnetic moments in the sample. The ZF-$\mu^{+}$SR spectra
were collected over the temperature range 5 - 250~K. The
ZF-$\mu$SR spectra are weakly damped with an increasing relaxation
when $T_\mathrm{C}$ is approached from above, see inset of
Fig.~\ref{Fig-mSR-highT}(a).

To analyze our experimental data, a stretched exponential form was
used at all temperatures above T$_c$:
\begin{equation}\label{P(t)aboveTc}
    P(t) = e^{-(\lambda t)^{\beta}}.
\end{equation}

Here, $\lambda$ represents the relaxation rate and $\beta$ the
exponent. The spin relaxation in all this temperature range is
attributed static nuclear moments and rapidly fluctuating cobalt
moments in the paramagnetic phase. This stretched exponential
function, Eq.~\ref{P(t)aboveTc}, has been found to be appropriate
in cases where a complex non-exponential fluctuation pattern is
expected, e.g. due to broad distribution of fluctuation times
\cite{campbell}. The temperature dependence of $\lambda(T)$ is
shown in Fig.~\ref{Fig-mSR-highT}(a). As the temperature
approaches the magnetic phase transition from above the magnetic
fluctuations slow down and lead to the sharp increase of
$\lambda(T)$ at $T_\mathrm{C}=160$~K for
La$_{0.5}$Ba$_{0.5}$CoO$_{2.94}$ and 180~K for
La$_{0.5}$Ba$_{0.5}$CoO$_{2.985}$, respectively. The temperature
dependence of the stretched exponential exponent $\beta$ above
$T_\mathrm{C}$ has been depicted in Fig.~\ref{Fig-mSR-highT}(b)
for both samples with $\delta$=~2.94 and $\delta$=~2.985,
respectively. In magnetically homogenous systems $\beta$ is equal
to unity whereas in inhomogenous systems with spin frustration
$\beta$ becomes smaller than unity \cite{keren}. The stretched
exponential exponent $\beta$ reaches a value of about
$\frac{1}{3}$ at $T_\mathrm{C}$ for both samples which is  typical
for inhomogeneous systems like spin or cluster glasses.

\begin{figure}[htbp]
\begin{center}
\center{\includegraphics[width=0.87\columnwidth,angle=0,clip]{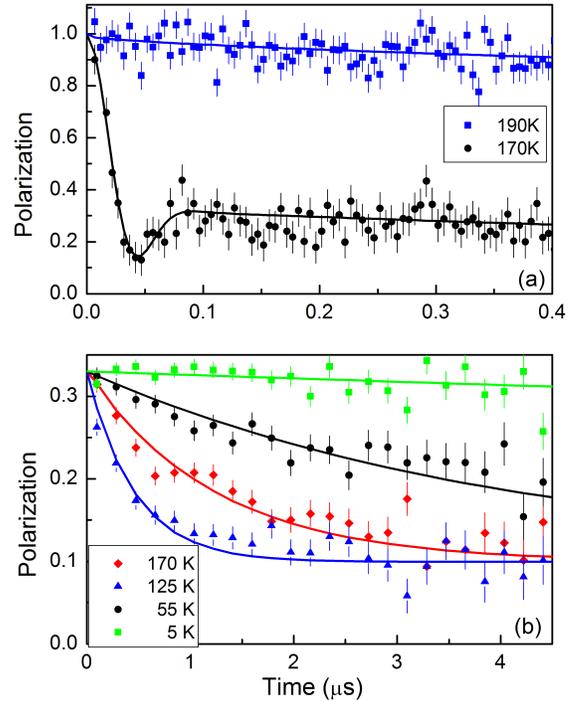}}
\caption{The zero-field muon spin relaxation spectra at various
temperatures for La$_{0.5}$Ba$_{0.5}$CoO$_{2.985}$. a) The early time spectra at
170~K, indicates that the muon depolarization is mostly due to
static fields. b) The damping of the $1/3$-tail of the late time
spectra reveals the presence of magnetic fluctuations in 70\% of the
sample volume well below $T_\mathrm{C}=180$~K.} \label{Fig-mSR-raw}
\end{center}
\end{figure}

Below $T_\mathrm{C}$ the ZF-$\mu$SR spectra change completely. In
Fig.~\ref{Fig-mSR-raw}(a) a spectrum at 170~K (below
$T_\mathrm{C}=180$~K) is compared to a spectrum at 190~K for
La$_{0.5}$Ba$_{0.5}$CoO$_{2.985}$.
A strong depolarization of $2/3$ of $P(t)$ is observed at early
times, indicating a wide distribution of internal magnetic fields
at the muon site, consistent with a magnetically inhomogeneous
system. Note that the early time relaxation is dominated by the
presence of the \emph{static component} of the field distribution
and that the depolarization of $2/3$ of the $\mu$SR signal implies
that 100\% of the sample is magnetic below $T_\mathrm{C}$. If the
field distribution at the muon site would be purely static within
the time window of $\mu$SR the fast relaxation of $2/3$ of the
signal at early times would be followed at late times by a
non-relaxing tail with $P(t)=~1/3$. This tail reflects, in a
powder average, the static field components which are longitudinal
to the initial muon spin which do not depolarize $P(t)$. This
behavior is not observed in the present samples of
La$_{0.5}$Ba$_{0.5}$CoO$_{2.985}$ and
La$_{0.5}$Ba$_{0.5}$CoO$_{2.94}$. The late time spectra of
La$_{0.5}$Ba$_{0.5}$CoO$_{2.985}$ for different temperatures below
$T_\mathrm{C}$ are shown in Fig.~\ref{Fig-mSR-raw}(b). Clearly, a
relaxation of the $1/3$-tail of the spectrum is observed for
temperatures well below $T_\mathrm{C}$. This relaxation is purely
due to fluctuating electronic moments, i.e. it is of dynamic
origin, see e.g. \cite{reotier04,klauss04}. The strongest dynamic
relaxation is observed at $T=~125$~K, i.e 55~K below
$T_\mathrm{C}$. Since $P(t)$ approaches 0.1 at late times it is
evident that only 70\% of the muons experience this dynamic
relaxation. In other words, a phase separation into a purely
static magnetic phase (30\%) and into a fraction of the sample
experiencing slow magnetic fluctuations (70\%) is observed below
$T_\mathrm{C}$. Below 125~K the dynamical relaxation decreases and
becomes insignificant at very low temperatures where the sample
becomes fully static at a temperature of 5~K.

Based on the qualitative results discussed above a fit
function given by the sum of a static and a dynamic depolarization function was used for the spectra below $T_\mathrm{C}$:
\begin{equation}\label{p(t)}
    P(t) = 0.3P_\mathrm{static} + 0.7P_\mathrm{dynamic}
\end{equation}
The two functions used are:
\begin{equation}\label{4}
P_\mathrm{static} =  \frac{2}{3}\cos(2\pi f t) e^{-\frac{1}{2}(\Delta
t)^{2}} + \frac{1}{3}
\end{equation}
\begin{equation}\label{4}
P_\mathrm{dynamic} =  [ \frac{2}{3}\cos(2\pi f t) e^{-\frac{1}{2}(\Delta
t)^{2}} + \frac{1}{3}]  e^{-\lambda t}
\end{equation}
Here, $\Delta$ measures the width of the static field
distribution, while the muon precession frequency $f$ is the mean
value of this distribution. $\lambda$ is the dynamic relaxation
rate. The $\mu$SR signal $P_\mathrm{dynamic}$ which is connected
to the magnetic phase in which static order and magnetic
fluctuations are simultaneously present is described by a product
of a static and a dynamic relaxation function. This estimation is
valid in the present case, where $f \gg \lambda$ \cite{hayano79}.
Static Lorentz and Gaussian Kubo-Toyabe functions have been tested
for the static relaxation function also, but only unsatisfactory
results have been obtained.
\begin{figure}[htbb]
\begin{center}
\center{\includegraphics[width=0.87\columnwidth,angle=0,clip]{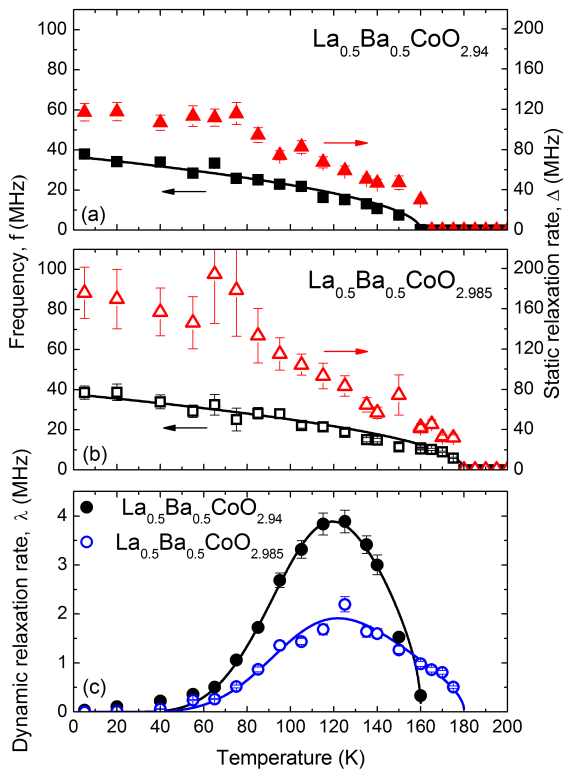}}
\caption{(a) and (b) show the temperature dependence of the static
relaxation rate $\triangle$(T) for
La$_{0.5}$Ba$_{0.5}$CoO$_{2.94}$ and
La$_{0.5}$Ba$_{0.5}$CoO$_{2.985}$ together with the $\mu$SR
frequency $f$ evolution. The solid lines in a) and b) are fits
according to Eq.~\ref{mean_field}. (c)~Temperature dependence of
the dynamical relaxation rate for the temperatures below T$_c$.
The lines represent the fit results according to the Redfield
theory.} \label{Fig-mSR-fit-results}
\end{center}
\end{figure}
Figures~\ref{Fig-mSR-fit-results}(a) and (b) display the fit
results of the static parameters of the field distribution for
La$_{0.5}$Ba$_{0.5}$CoO$_{2.94}$ and
La$_{0.5}$Ba$_{0.5}$CoO$_{2.985}$ as a function of temperature,
respectively. $\triangle (T)$ and the $\mu$SR frequency $f$
develop  below $T_\mathrm{C}$ in the ferromagnetic phase and
continuously increase decreasing the temperature. The dynamic
relaxation rate $\lambda~(T)$ for both compounds is shown in
Fig.~\ref{Fig-mSR-fit-results}(c). It displays a maximum at
$T_\mathrm{f}=125$~K for both doping levels. Roughly speaking,
this behavior reflects the slowing down of magnetic fluctuation
with the fluctuation time $\tau$ crossing the time window of the
$\mu$SR technique. Below $T_\mathrm{f}$ the fluctuations freezes
out gradually and finally $\lambda$ approaches zero at very low
temperatures. A possible explanation for this behavior is a
freezing of local charge and/or spin state fluctuations between
neighboring Co$^{3+}$ and Co$^{4+}$ ions which freeze below
$T_\mathrm{f}$. The idea of this kind of freezing is supported by
the fact that no anomaly is observed in the macroscopic
measurements, see Fig.~\ref{Fig-Magn}. The dynamic relaxation rate
$\lambda$ was evaluated using a modified Redfield theory
\cite{slichter}. According to Redfield theory the dynamic
relaxation rate
\begin{equation} \label{redfield}
\lambda = 2\sigma^2 \frac{\nu_\mathrm{c}}{\nu_\mathrm{c}^2 + \omega_\mu^2}
\end{equation}
is given by the fluctuation rate $\nu_\mathrm{c} = 1/\tau$, the
muon spin Larmor precession frequency $\omega_\mu = 2\pi f$, and
$\sigma$ describing the static field distribution. This formula is
valid for a isotropic field distribution in the fast fluctuation
regime ($\nu_\mathrm{c} >> \sigma$). Following the argumentation
of Baabe \emph{et al.} \cite{baabe04} the Redfied theory modifies
to
\begin{equation} \label{mod_redfield}
\lambda = 1.5\sigma_\mathrm{VV}^2 \frac{\nu_\mathrm{c}}{\nu_\mathrm{c}^2 + \omega_\mu^2}
\end{equation}
when electronic spins are the reason for the field fluctuations at
the muon site, which themselves precess fast in a static magnetic
field given by a magnetic environment.  $\sigma_\mathrm{VV}$ is
the static Van Vleck linewidth. The solid lines in
Fig.~\ref{Fig-mSR-fit-results}(c) are fits of the dynamic
relaxation rate using the modified Redfield theory
Eq.~\ref{mod_redfield}. To evaluate the function $\lambda (T)$ the
temperature dependences of $\omega_\mu (T)$ and
$\sigma_\mathrm{VV} (T)$ have to be known. From a fit of the
measured $\mu$SR frequencies $f(T)$ to a simple mean field
approach, $\omega_\mu (T)$ is obtained (see solid lines in
Fig.~\ref{Fig-mSR-fit-results}(a) and (b)):
\begin{equation} \label{mean_field}
\omega_\mu (T) = 2\pi f_0 (1 - T/T_\mathrm{C})^{0.5}
\end{equation}
The relation between $\sigma_\mathrm{VV}$ and $\omega_\mu$ has
been determined by evaluating the measured dynamic relaxation rate
$\lambda$ at its maximum value, where, according to Eq.~\ref{mod_redfield}, $\nu_\mathrm{c} =
\omega_\mu$ and the following relation holds:
\begin{equation} \label{sigma}
\lambda_\mathrm{max} = \frac{3}{4} \frac{\sigma_\mathrm{VV}^2}{\omega_\mu}
\end{equation}
In other words, $\omega_\mu (T)$ and the proportionality constant
between $\sigma_\mathrm{VV}$ and $\omega_\mu$ have been determined
experimentally. Strictly speaking the proportionality constant
$4/3 \lambda_\mathrm{max}$ has been measured at the maximum of
$\lambda(T)$ only. The analysis of $\lambda (T)$ using the
Redfield formalism has been performed assuming that this
proportionality holds for all temperatures, i.e. :
\begin{equation} \label{sigma}
\lambda_\mathrm{max} = \frac{3}{4} \frac{\sigma_\mathrm{VV}^2 (T)}{\omega_\mu (T)}
\end{equation}
Having the other parameters determined independently, the only
free parameter in the modified Redfield formula
Eq.~\ref{mod_redfield} is the magnetic fluctuation rate
$\nu_\mathrm{c}$. The data shown in
Fig.~\ref{Fig-mSR-fit-results}~c) are well described by an
activated behavior of the fluctuation rate
\begin{equation} \label{activated}
\nu_\mathrm{c} = \nu_0 \exp(- E_\mathrm{a}/T)
\end{equation}
with an attempt frequency $\nu_0$ and an activation energy
$E_\mathrm{a}$, the obtained fit results are summarized in
Tab.~\ref{table-arrhenius}.
 \begin{table}
 \caption{Attempt frequency $\nu_0$ and activation energy
$E_\mathrm{a}$ obtained from the fit shown in
Fig~\ref{Fig-mSR-fit-results}~c) using the modified Redfield
theory explained in the text.\label{table-arrhenius}}
 \begin{ruledtabular}
 \begin{tabular}{l c c}
& La$_{0.5}$Ba$_{0.5}$CoO$_{2.94}$ & La$_{0.5}$Ba$_{0.5}$CoO$_{2.985}$ \\
$\nu_0$ (MHz)   & 1952(234)& 2241(526) \\
$E_\mathrm{a}$ (K) & 336(14) & 343(28)\\
 \end{tabular}
 \end{ruledtabular}
 \end{table}
The fit to the data is excellent over the full temperature range
even though the described formalism is strictly speaking only
valid at high temperatures and slightly below the maximum of
$\lambda(T)$. At low temperatures the fluctuations are so slow,
that the Redfield formula can not be applied anymore. For both
compounds, an activation energy of $E_\mathrm{a}\approx 340$~K is
observed, which is a typical activation energy for a Jahn-Teller
like phonon. The present quantitative data are in good agreement
with the neutron data published by Fauth \emph{et al.}
\cite{fauth} where a cooperative Jahn-Teller deformation develops
lowering the temperature below 140~K where the metal-insulator
transition takes place. Therefore we propose that the freezing of
spin components at $T_\mathrm{f}$ has its origin in the
Jahn-Teller distortion and/or orbital ordering. In other words, we
suggest that the observed magnetic fluctuations stem from local
charge and corresponding spin state fluctuations between the
Co$^{3+}$ and Co$^{4+}$ ions which are driven by Jahn-Teller
fluctuations. Due to the La/Ba-site random disorder, the CoO$_6$
octahedra will be distorted locally which additionally favors the
localization of the conduction electrons and an insulator state at
lower temperatures. The charge carrier localization accompanying
the Jahn-Teller distortions probably shifts the balance of the
competing double and super exchange towards the AFM superexchange
and in turn the randomly distributed ferromagnetic -
antiferromagnetic interactions lead to a magnetic glassy state
observed below $T_\mathrm{f}$.

\section{Conclusion}
In conclusion, susceptibility and $\mu$SR measurements on two
La$_{0.5}$Ba$_{0.5}$CoO$_{2.5+\delta}$ samples with different
oxygen content, $\delta$ =~0.44 and 0.485 have been performed in
order to study its static and dynamic magnetic properties. Both
compounds are paramagnetic at high temperatures and a transition
to an inhomogeneous ferromagnetic phase is observed at
$T_\mathrm{C}=160$~K and 180~K, respectively. The samples are
electronically phase separated into phases possessing static
magnetic order (30\%) and a major sample volume showing slow
magnetic dynamics of electronic moments. A maxima in the dynamic
magnetic relaxation rate at $T_\mathrm{f}=125$~K signals a
freezing of spin components well below $T_C$. The simultaneous
occurrence of the spin freezing and static cooperative Jahn-Teller
distortions below $T_\mathrm{f}$ and the observation that the
magnetic fluctuations are thermally activated with a typically
activation energy of Jahn-Teller like phonons suggest that the
magnetoelastic coupling is important to the understanding of the
low-temperature spin dynamics of
La$_{0.5}$Ba$_{0.5}$CoO$_{2.5+\delta}$ and that it is responsible
for the observed spin freezing. We propose that the magnetic
fluctuations originate from local charge and hence spin state
fluctuations of neighboring Co$^{3+}$ and Co$^{4+}$ ions driven by
dynamic Jahn-Teller distortions. This interpretation is also
consistent with the observed absence of a change of the
macroscopic magnetization at $T_\mathrm{f}$. Below $T_\mathrm{f}$
a magnetically glassy state, i.e. a re-entrant spin glass phase is
entered as evidenced by the difference of FC and ZFC magnetization
curves and a typical frequency dependence of the
ac-susceptibility.

\begin{acknowledgments}
This work was performed at the Swiss Muon Source, Paul Scherrer
Institut, Villigen, Switzerland with the financial support from
Swiss National Science Foundation through grant N$\rm ^o$
200021-101634 and NCCR MaNEP.
\end{acknowledgments}

\end{document}